\documentclass[epj]{webofc}
\usepackage[utf8]{inputenc}
\usepackage[varg]{txfonts}
\usepackage{booktabs}
\usepackage{xcolor}
\definecolor{darkred}{rgb}{0.4,0.0,0.0}
\definecolor{darkgreen}{rgb}{0.0,0.4,0.0}
\definecolor{darkblue}{rgb}{0.0,0.0,0.4}
\usepackage[bookmarks,linktocpage,colorlinks,
    linkcolor = darkred,
    urlcolor  = darkblue,
    citecolor = darkgreen]{hyperref}
\usepackage{subfigure}
\wocname{EPJ Web of Conferences}
\woctitle{Lattice2017}

\newcommand{\be}{\begin{equation}}
\newcommand{\ee}{\end{equation}}

\newcommand{\bea}{\begin{eqnarray}}
\newcommand{\eea}{\end{eqnarray}} 

\newcommand{\la}{\langle}
\newcommand{\ra}{\rangle}

\newcommand{\Z}{\mathbb{Z}}
\newcommand{\R}{\mathbb{R}}
\newcommand{\N}{{\kern+.25em\sf{N}\kern-.78em\sf{I} \kern+.78em\kern-.25em}}

\newcommand{\gtapprox}{\raisebox{-0.5ex}{$\,\stackrel{>}{\scriptstyle\sim}\,$}}
\newcommand{\ltapprox}{\raisebox{-0.5ex}{$\,\stackrel{<}{\scriptstyle\sim}\,$}}

\begin{document}
%%%%%%%%%%%%%%%%%%%%%%%%%%%%%%%%%%%%%%%%%%%%%%%%%%%%%%%%%%%%%%%%%%%%%%%%%%%%%
\selectlanguage{english}
%----------------------------------------------------------------------------
\title{%
Topological Susceptibility under Gradient Flow}
%----------------------------------------------------------------------------
\author{%
\firstname{H\'{e}ctor} \lastname{Mej\'{\i}a-D\'{\i}az}\inst{1}\fnsep\thanks{Speaker, \email{he_mejia@yahoo.com.mx}} \and
\firstname{Wolfgang} \lastname{Bietenholz}\inst{1} \and
\firstname{Krzysztof} \lastname{Cichy}\inst{2,3} \and
\firstname{Philippe} \lastname{de Forcrand}\inst{4,5} \and \\
\firstname{Arthur} \lastname{Dromard}\inst{6} \and
\firstname{Urs} \lastname{Gerber}\inst{1,7} \and
\firstname{Ilya Orson} \lastname{Sandoval}\inst{1}
}
%----------------------------------------------------------------------------
\institute{%
Instituto de Ciencias Nucleares,
Universidad Nacional Aut\'{o}noma de M\'{e}xico \\
\hspace*{1mm} A.P.\ 70-543, C.P.\ 04510 Ciudad de M\'{e}xico, Mexico
\and
Goethe-Universit\"{a}t Frankfurt am Main, Institut für Theoretische
Physik \\ \
\hspace*{1mm} Max-von-Laue-Stra\ss e 1, 60438 Frankfurt am Main, Germany
\and
Faculty of Physics, Adam Mickiewicz University,
Umultowska 85, 61-614 Pozn\'{a}n, Poland
\and
Institut f\"{u}r Theoretische Physik, ETH Z\"{u}rich,
Wolfgang-Pauli-Strasse 27, CH--8093 Z\"{u}rich, Switzerland
\and
CERN, Theory Division, CH-1211 Gen\`{e}ve 23, Switzerland
\and
Institut f\"{u}r Theoretische Physik, Universit\"{a}t 
Regensburg, D-93040 Regensburg, Germany
\and
Instituto de F\'{\i}sica y Matem\'{a}ticas,
Universidad Michoacana de San Nicol\'{a}s de Hidalgo \\
\hspace*{1mm} Edificio C-3, Apdo.\ Postal 2-82, C.P.\ 58040,
Morelia, Michoac\'{a}n, Mexico
}
%----------------------------------------------------------------------------
\abstract{%
We study the impact of the Gradient Flow on the topology in
various models of lattice field theory.
The topological susceptibility $\chi_{\rm t}$ is
measured directly, and by the {\em slab method}, which is
based on the topological content of sub-volumes (``slabs'')
and estimates $\chi_{\rm t}$ even when the system
remains trapped in a fixed topological sector.
The results obtained by both methods are essentially consistent,
but the impact of the Gradient Flow on the characteristic quantity
of the slab method seems to be different in 2-flavour QCD and in
the 2d O(3) model. In the latter model, we further address
the question whether or not the Gradient Flow leads to a finite
continuum limit of the topological susceptibility (rescaled by
the correlation length squared, $\xi^{2}$). This ongoing study is
based on direct measurements of $\chi_{\rm t}$ in $L \times L$
lattices, at $L/\xi \simeq 6$.}

\maketitle

\section{Introduction}\label{intro}

In some quantum field theories, the set of configurations is divided
into topological sectors, labelled by a topological charge $Q \in \Z$.
This is the case in QCD, and in $N$-dimensional O($N+1$)
models (with periodic boundary conditions for the gluon and spin fields),
due to $\Pi_{4}[{\rm SU}(3)] = \Z$ and $\Pi_{N}[S^{N}] = \Z$.
Hence this class of models includes 2-flavour QCD, as well as the
1d O(2) and the 2d O(3) model, which we are going to deal with.

For usual lattice actions, all configurations can be continuously
deformed into one another, at finite action, hence there are no
topological sectors in a strict sense. Exceptions are {\em topological
  lattice actions,} with a sharp cutoff for the angles between nearest
neighbour spin variables \cite{topact}, or for each plaquette variable
\cite{PdFU1} (see also Ref.\ \cite{Luscher:plaqact})
in spin models and gauge theories,
respectively. However, even for conventional lattice actions there
are established ways to divide the configurations into sectors, which
turn into topological sectors in the continuum limit.

Here we consider 2-flavour QCD with twisted-mass quarks \cite{twist}
(at full twist) and the Wilson gauge action. For the O($N$) models
we employ the standard lattice action,
\be
S[\vec e \, ] =
\beta \sum_{\la xy \ra} (1 - \vec e_{x} \cdot \vec e_{y}) \ ,
\quad \vec e_{x} \in S^{N-1} \ \ \forall x \ ,
\quad N = 2 ~ {\rm or} ~ 3 \ ,
\ee
where the sum runs over all nearest neighbour lattice
sites. For these O($N$) models we apply the geometric definition of
the topological charge density on the lattice \cite{BergLuscher},
which leads to integer charges $Q \in \Z$.
In QCD we use a clover discretisation of $F_{\mu \nu} \tilde F_{\mu \nu}$,
where $F$ is the field strength tensor.

In all cases under consideration, parity symmetry implies
$\la Q \ra =0$, hence the topological susceptibility takes the form
\be
\chi_{\rm t} = \frac{1}{V} \la Q^{2} \ra \ , \quad
Q~:~{\rm topological~charge},\ V~:~{\rm volume}.
\ee

\section{The slab method to measure the topological susceptibility
$\chi_{\rm t}$}\label{slab}

Once we have fixed a formulation of the topological charge on the lattice,
it is straightforward to measure $\chi_{\rm t}$ by means of Monte Carlo
simulations, {\em if} the Markov chain frequently changes $Q$,
such that the sectors are sampled correctly. In practice, however, such
simulations are often confronted with the severe problem of ``topological
freezing'': in particular, the algorithms, which proceed in small update
steps, tend to get stuck in one topological sector for a huge number
of steps, since the topological sectors are effectively separated by
high potential barriers. The autocorrelation time with respect to $Q$
increases with a high power of the inverse lattice spacing as we
approach the continuum limit (``topological slowing down''),
see {\it e.g.}\ Ref.\ \cite{Rainer}.

A variety of approaches to handle this problem is reviewed in
Ref.\ \cite{Endres}. One strategy aims at extracting physical
observables even from a Markov chain which is entirely trapped in
a single topological sector. For general observables such a method
was suggested in Ref.\ \cite{BCNW}, and tested and
extended in Refs.\ \cite{Schwing,BCNWt1,BCNWt2,Arthur}. More
specifically, a procedure to measure $\chi_{\rm t}$ within a fixed
topological sector was proposed in Ref.\ \cite{AFHO} and tested in
Refs.\ \cite{AFHOt1,Schwing,AFHOt2,Arthur}. Here we consider the
{\em slab method} as another way to evaluate $\chi_{\rm t}$ from
data obtained at fixed $Q$ (actually data from $\pm Q$ can be combined).
The idea was mentioned in Ref.\ \cite{PdF99},
implemented in Ref.\ \cite{slab}, and further
explored in Refs.\ \cite{slabproc1,Arthur,Lat16}.
A different variant was applied in Ref.\ \cite{slabjap}, and
there are similarities with the approach in Ref.\ \cite{LSD14}.

We briefly review the simplest version of the slab method, which
assumes the statistical distribution of the topological charges
to be Gaussian \cite{slab},
$p(Q) \propto \exp ( -Q^{2}/(2 \chi_{\rm t}V))$.
We split the volume $V$ into two sub-volumes (``slabs'')
$xV$ and $(1-x)V$ \ ($0<x<1$). By summing up the topological charge
density in each of them, in a configuration of topological charge
$Q$, we obtain the slab charges $q, \, Q-q \in \R$ (they do not
need to be integer, since the slabs do not have periodic boundaries).
At fixed $x$, $V$ and $Q$, the corresponding slab probability
distributions $p_{1}$ and $p_{2}$ obey
\be
p_{1}(q) \, p_{2}(Q-q) \propto \exp \left( - \frac{1}{2 \chi_{\rm t}V}
\frac{q{'}^{2}}{x(1-x)} \right) \ , \quad q' = q - xQ \ .
\ee
Measuring $\la q^{2} \ra$ yields a value for
$\la q{'}^{2} \ra = \la q^{2} \ra - x^{2} Q^{2}$. A sequence of such
measurements, at different parameters $x$, enables a fit to the
prediction
\be \label{predict}
\la q{'}^{2} \ra = \chi_{\rm t} V \, x (1-x) \ ,
\ee
which provides a result for $\chi_{\rm t}$. In practice, the most
reliable fitting regime is around the center, $x \approx 0.5$,
because the size of the slabs should be large compared to that
of topological excitations; one should not include $x \gtapprox 0$
or  $x \ltapprox 1$ (these regions involve very small slabs, where
the Gaussian distribution is not a good approximation).

\subsection{Results by the slab method under Gradient Flow (GF)}\label{GF}

As a smoothing procedure for lattice field configurations, the GF
corresponds to a renormalisation group scheme. When the GF proceeds,
the distinction between topological sectors becomes more marked,
approaching the continuum feature of a total separation
\cite{LuscherGF1,LuscherGF2}.

The slab method has been tested in 2-flavour QCD, with
twisted-mass quarks and the Wilson gauge action, in a volume
$16^{3} \times 32$, at $\beta = 3.9$ and bare mass $0.015$, which
corresponds to a pion mass of $m_{\pi} \simeq 650 \, {\rm MeV}$
and a lattice spacing $a \simeq 0.079 \, {\rm fm}$ \cite{slabproc1,Lat16}.

\begin{figure}[thb]
  \centering
  \subfigure
    {\includegraphics[width=7cm,clip]{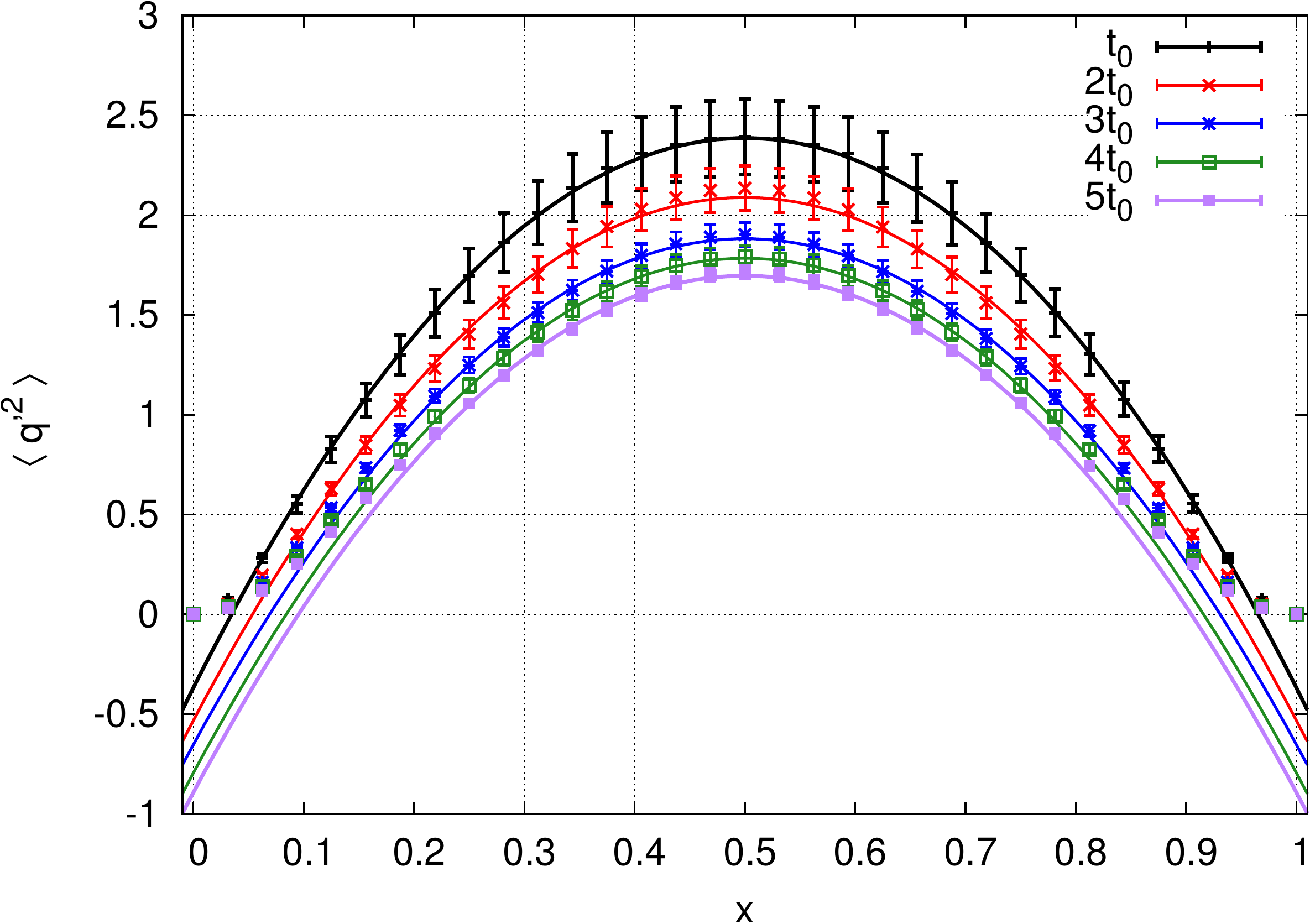}}\hfill  
  \subfigure
    {\includegraphics[width=7.1cm,clip]{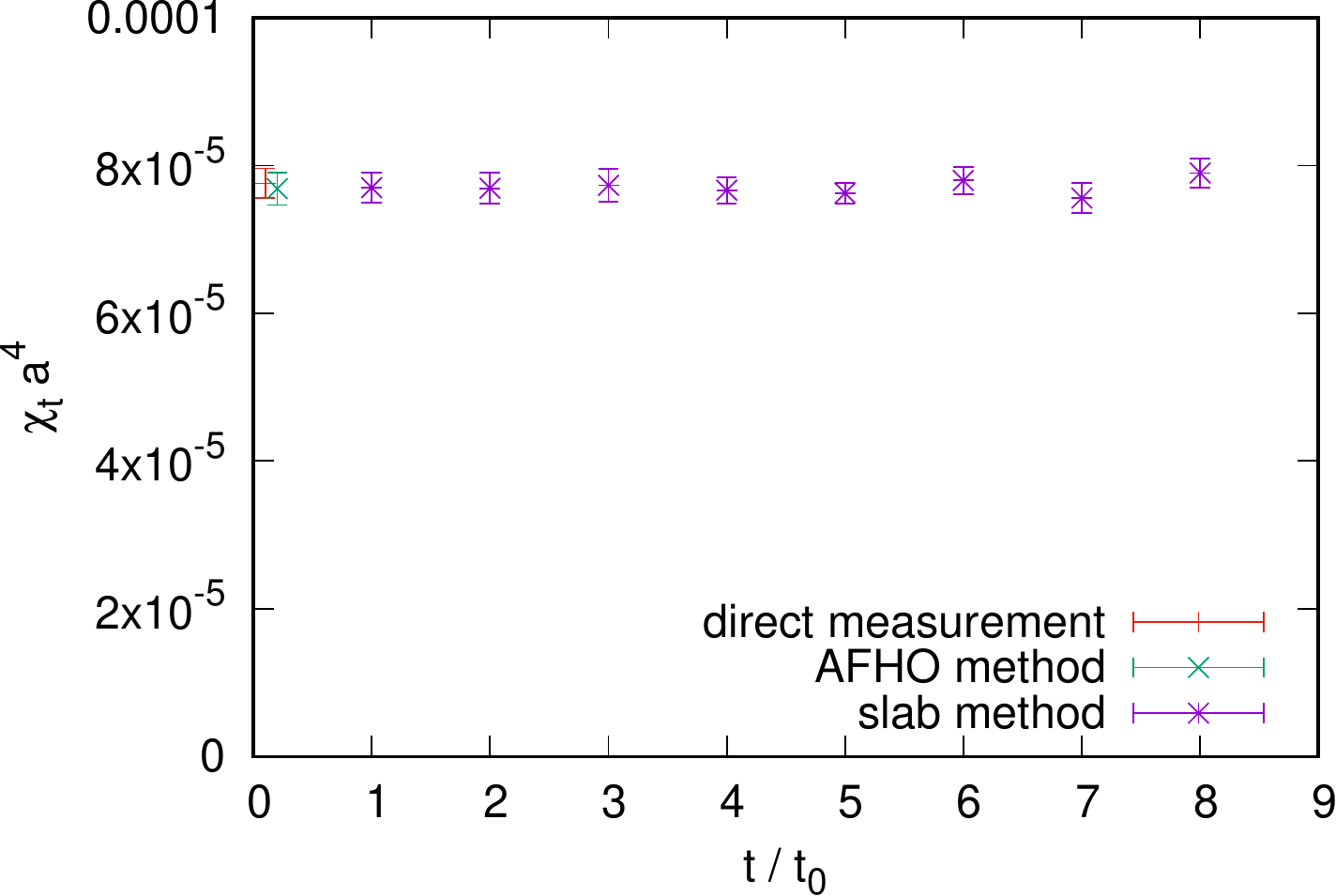}} 
    \vspace{-2mm}
    \caption{On the left: the expectation value $\la q{'}^{2} \ra$,
      as a function of the parameter $x = V_{\rm slab}/V$, in 2-flavour
      QCD, in the sector $|Q|=1$.
      Combined fits to eq.\ (\ref{extend}), in the sectors $|Q|\leq 2$,
      at different instances of the GF time,
      lead to the results for the topological susceptibility
      $\chi_{\rm t}$ in the plot on the right. They agree with a direct
      measurement, and with the method of Ref.\ \cite{AFHO} (AFHO method).}
    \label{QCDslab}
     \vspace{-3mm}
\end{figure}

The GF was implemented with the Runge-Kutta method; the results
with time step $dt = 0.01$ and $0.001$ agree.
The GF flow time unit was fixed to $t_{0}/a^{2} = 2.42$, based on the
criterion proposed in Refs.\ \cite{LuscherGF1,LuscherGF2}:
$\la E \ra t_{0} =0.3$, where $\la E \ra$ is the mean energy density.
The effect of the GF on the curves to be fitted in the slab method
is shown in Figure \ref{QCDslab} (left). As the GF proceeds, the fit
has to be restricted to a narrower interval centered at $x=0.5$.
Moreover, the fitting function (\ref{predict}) has to be extended to
\be \label{extend}
\la q{'}^{2} \ra = \chi_{\rm t} V \, x (1-x) - c \ ,
\ee
where $c$ is a constant (with respect to $x$), which increases
roughly like $\simeq 0.38 \sqrt{t}$ \ \cite{Lat16}.
The fits at different instances of the GF time,
$t = t_{0}, \, 2 t_{0} \dots 8t_{0}$, yield very stable results
for $\chi_{\rm t}$, which are compatible both with a direct measurement
($\chi_{\rm t} a^{4} = 7.8(2) \cdot 10^{-5})$,
and with the method of Ref.\ \cite{AFHO}
($\chi_{\rm t} a^{4} = 7.7(2) \cdot 10^{-5})$, see Figure \ref{QCDslab}
(right). In the interval $t = t_{0} \dots 4 t_{0}$ we consistently
obtain $\chi_{\rm t} a^{4} = 7.7(2) \cdot 10^{-5}$. This is a
success of the slab method, but the r\^{o}le of the subtractive
constant $c$ is not obvious.\\

In the continuum, the Gradient Flow in O($N$) models takes the form
\cite{MakSuz}
\be
\partial_{t} e(t,x)^{i} = \sum_{j=1}^{N} P^{ij}(t,x) \ \Delta e(t,x)^{j} \ ,
\quad P^{ij}(t,x) = \delta^{ij} - e(t,x)^{i} e(t,x)^{j} \ ,
\ee
where $t \geq 0$ is the GF time (of dimension [length]$^{2}$)
and $\Delta$ is the Laplace operator, which we handle by standard
lattice discretisation. In order to proceed in discrete flow time
steps, we apply the Runge-Kutta method. For a given configuration
we first compute the gradients of all spin variables (at all times
required by the Runge-Kutta 4-point scheme), then all spins are
modified simultaneously, with a time step $dt = 10^{-4}$; afterwards
the spins are normalised again.

\begin{figure}[tp]
   \centering
   \subfigure
  {\includegraphics[width=7cm,clip]{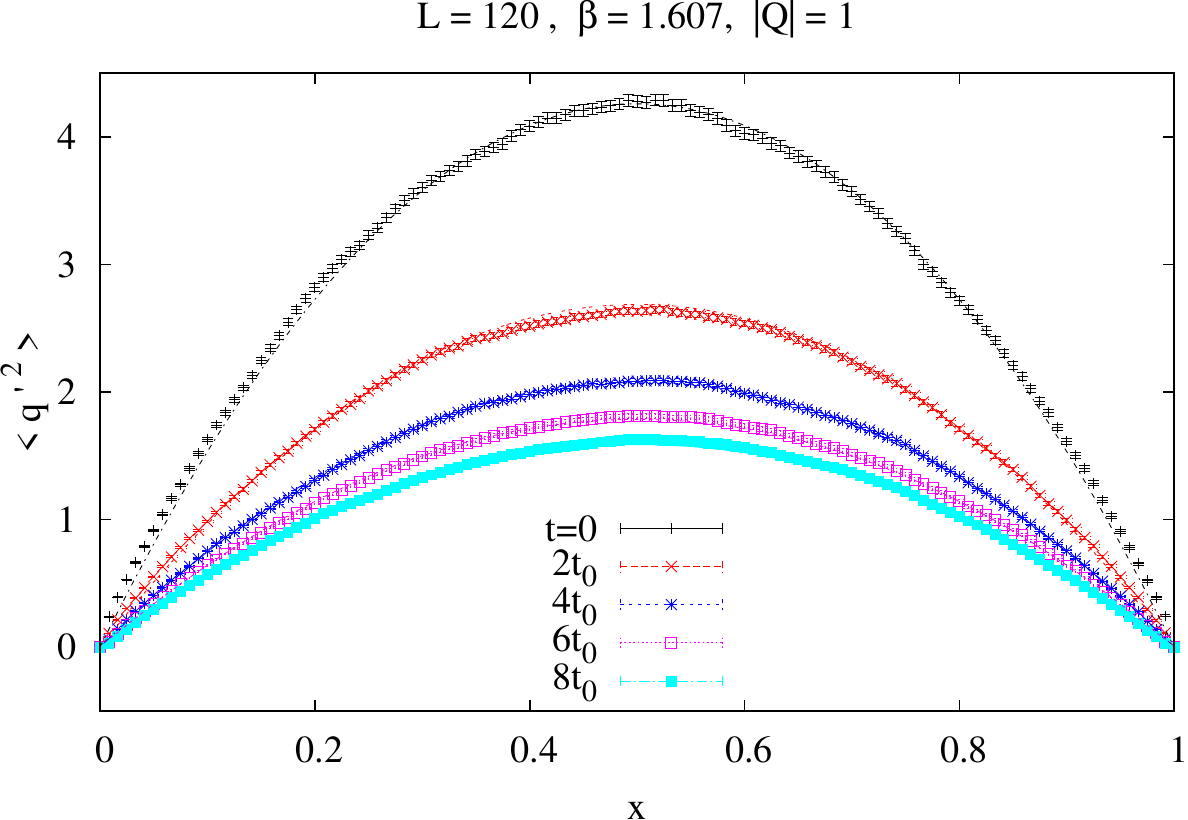}}\hfill 
   \subfigure
 {\includegraphics[width=7cm,clip]{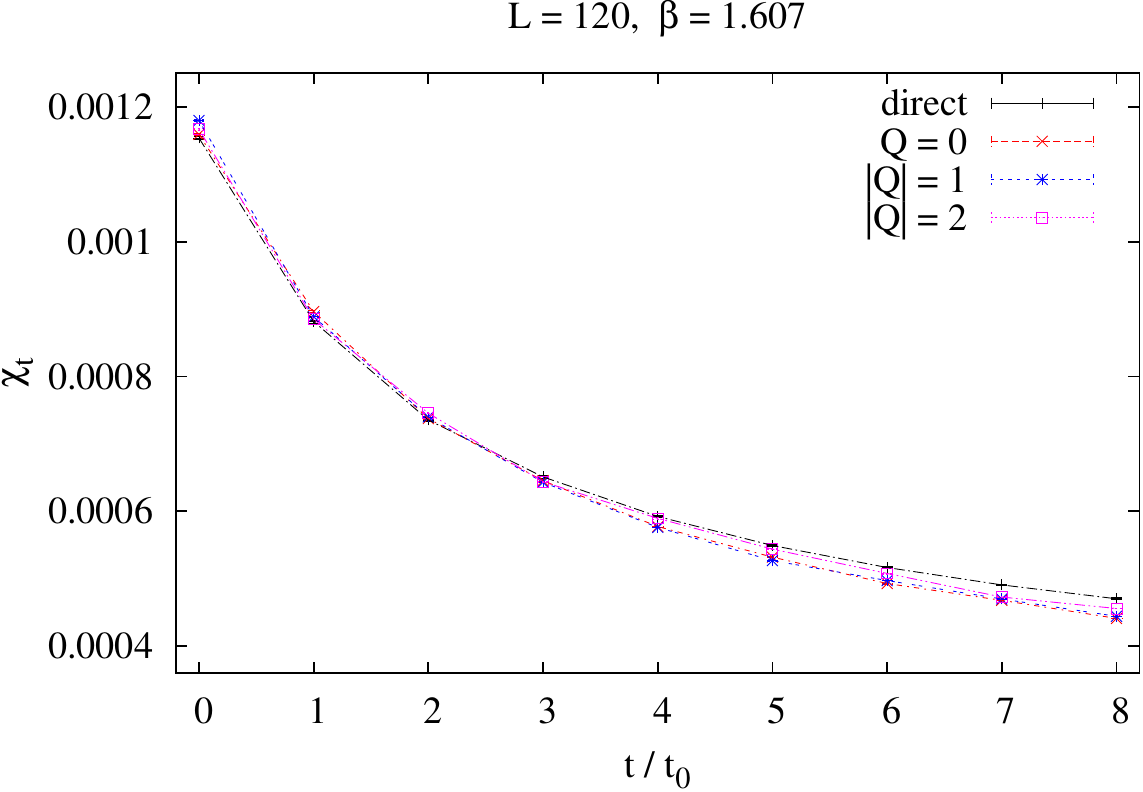}}
 \vspace{-2mm}
 \caption{On the left: the quantity $\la q{'}^{2} \ra (x)$, in the 2d
   O(3) model, in the sector $|Q|=1$, with and without GF.
   Fits to eq.\ (\ref{predict}), in the sectors $|Q|=0,\, 1,\, 2$,
   yield the results for $\chi_{\rm t}$, which are shown on the
   right. They are close to each other, and to the values from
   direct measurement (the GF drives them slightly apart).}
\label{2dO3slab}
   \vspace{-5mm}
\end{figure}

In this manner, we considered $L \times L$ lattices, for instance
with $L=120$, $\beta = 1.607$, where the correlation length amounts
to $\xi = 19.1(2)$ \cite{RA17}. Figure \ref{2dO3slab}
(left) shows the (approximate) slab parabolae obtained
for $\la q{'}^{2} \ra (x)$ in the sector $|Q|=1$,
in even multiples of the flow time unit $t_{0} = 0.0772$
(which obeys $\la E \ra t_{0} = 0.08$, cf.\ Section \ref{2dO3}).
We see a qualitative difference from the QCD result in
Figure \ref{QCDslab}: here the fits do not require any
subtractive constant, {\it i.e.}\ the
original formula (\ref{predict}) can be used, and
$\chi_{\rm t}$ keeps on decreasing as the GF proceeds
(the curvature of the parabola is reduced).
This feature was also observed in all data sets to be
reported in Section \ref{2dO3}. 
The fitting results for $\chi_{\rm t}$ --- obtained by the slab
method, separately in the sectors $|Q| = 0,\, 1,\, 2$ --- are
close to the directly measured values, as we see in Figure
\ref{2dO3slab} (right).
In this case, the direct measurement is not problematic, since our
simulations were carried out with the Wolff cluster algorithm,
which proceeds in non-local update steps, thus suppressing the
effect of topological freezing \cite{Wolff89}.\\

\begin{figure}[tp]
   \centering
   \subfigure
 {\includegraphics[width=7cm,clip]{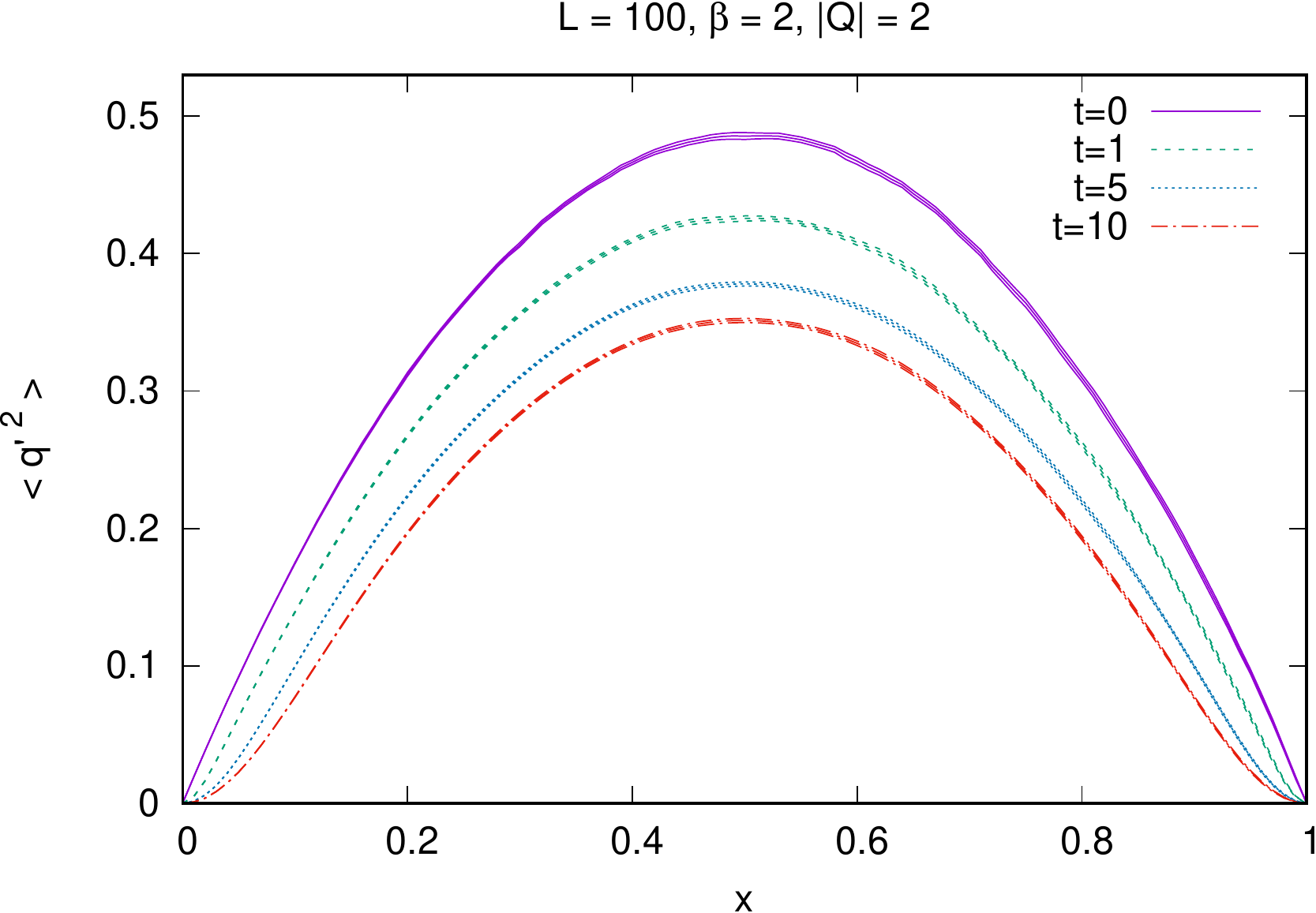}}\hfill
   \subfigure
 {\includegraphics[width=7cm,clip]{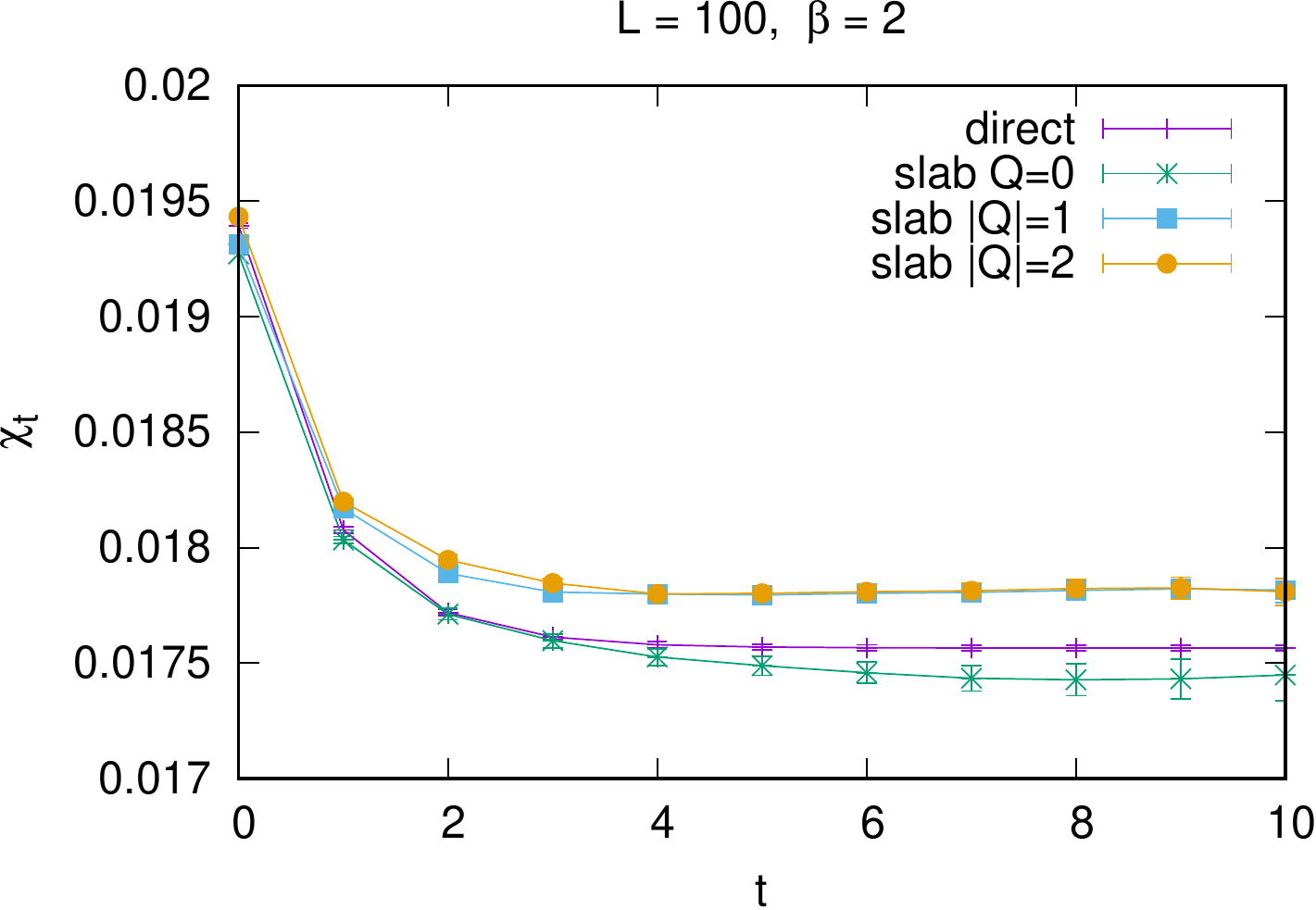}}
 \caption{The slab method in the 1d O(2) model: at short GF times,
   $\la q{'}^{2} \ra (x)$ (on the left) can be fitted to eq.\
   (\ref{predict}), but long flow times require an extension to
   eq.\ (\ref{extend}), and the exclusion of $x \gtapprox 0$ and
   $x \ltapprox 1$ from the fit. After a long flow time, the
   sector $Q=0$ provides the best results for $\chi_{\rm t}$,
   as the plot on the right shows.}
\label{1dO2slab}
\end{figure}

In order to investigate further these qualitatively different
behaviours, we tested the 1d O(2) model, or quantum rotor,
as a toy model. Here we refer to $\beta =2$, as an example.
In infinite volume we can compute analytically \cite{rot97}
(still for the standard lattice action, in lattice units)
\be
\beta = 2 ~: \quad \xi \simeq 2.779 \ ,
\quad \chi_{\rm t} \simeq 0.01936 \ ,
\ee
in agreement with our simulation results for size $L=100$
(we used again the cluster algorithm).
Regarding the slab method under GF, Figure \ref{1dO2slab} shows
the results for $\la q{'}^{2} \ra (x)$ in the sector $|Q|=2$
at various flow times $t$, and $\chi_{\rm t}$ as 
a function of $t$. For the scaling quantity we obtain numerically at
$t = 0 :~ \chi_{\rm t} \xi = 0.05388(6)$, and at
$t = 10 :~ \chi_{\rm t} \xi = 0.0496(1)$;
thus we gradually approach the continuum value of
$\chi_{\rm t} \xi = 1/(2 \pi^{2}) \simeq 0.05066$.
Regarding the different features in QCD and in the 2d O(3) model,
the quantum rotor at short flow times ($t = {\cal O}(1)$) seems
compatible with the latter, but at $t = {\cal O}(10)$ a non-negligible
constant has to be subtracted for a successful fit; at $t=10$ it
takes values $c \approx 0.08 \dots 0.09$ in the sectors $|Q| \leq 2$.

In the 2d O(3) model our flow times are short so far,
cf.\ Section \ref{2dO3}; the question whether the same
behaviour sets in after a long GF is under investigation.

\section{Topological scaling in the 2d O(3) model}\label{2dO3}

In a $d$-dimensional quantum field theory with topological sectors,
the dimensionless term $\chi_{\rm t}\, \xi^{d}$ is supposed to be
a scaling quantity, which converges to a finite value in the continuum
limit. For the 1d O(2) model
the continuum value $\chi_{\rm t}\, \xi = 1/(2 \pi^{2})$ is attained
without problems \cite{topact,BCNWt2,rot97}.
In QCD, and in SU($N$) Yang-Mills theories ($N\geq 2$),
a straight approach based on the expression
$\chi_{\rm t} = \sum_{x} \la q_{0} q_{x} \ra$, where $q_{x}$ is the
lattice topological charge density, faces problems. In conventional
formulations one encounters a divergence due to the point $x=0$,
which prevents the continuum scaling. In these cases, there are
known solutions to this problem, in particular the application
of the GF \cite{LuscherGF1,LuscherGF2}. 

In the 2d O(3) model, this question has been controversial in the
1980s and 1990s. The consensus is now that the quantity
$\chi_{\rm t}\, \xi^{2}$ seems to diverge in the continuum limit,
as first observed in \cite{BergLuscher}, and later underpinned
by semi-classical studies, {\it e.g.}\ in Refs.\ \cite{FarPap,MicSpen}.
Again the problem can be traced back to the topological density
correlation at distance zero (see {\it e.g.}\ Ref.\ \cite{topact}),
and the semi-classical picture suggests an abundance of very small
topological windings, so-called {\em dislocations},
as the continuum limit is
approached. Ref.\ \cite{Blatter} constructed and applied a
sophisticated (truncated) classically perfect lattice action,
with a host of couplings beyond nearest neighbour lattice sites,
which suppress such dislocations. Nevertheless, the simulation
results with this action still suggest a logarithmic divergence
of the term $\chi_{\rm t}\, \xi^{d}$ in the continuum limit
($\xi \to \infty$ in lattice units). Hence the continuum limit
of this popular model is generally assumed to be ill-defined, at least
with respect to its topology (the reason is again the contribution
$\la q_{0} q_{0}\ra$, see {\it e.g.}\
Refs.\ \cite{BalogPRD,BalogNPB,topact}).

However, the question remains whether or not this divergence could be
overcome by the GF; we gave preliminary results in
Ref.\ \cite{RA17}. In the following we summarise the status of
this study. So far it involves nine $L\times L$ lattices, in the range
of $L = 24 \dots 404$, where in each volume $\beta$ has been tuned
such that $L/\xi \simeq 6$. Therefore, increasing $L$ corresponds
to a controlled step towards the continuum limit, at a fixed and
large physical box size. The statistics in each volume are $10^{5}$
configurations (generated by the Wolff cluster algorithm, both
in the single-cluster and the multi-cluster version).
We repeat that we perform the GF with the Runge-Kutta 4-point
method, with a time step of $dt = 10^{-4}$, which is simultaneously
applied to all spin variables.

\begin{figure}[tp]
   \centering
   \subfigure
   {\includegraphics[width=7cm,clip]{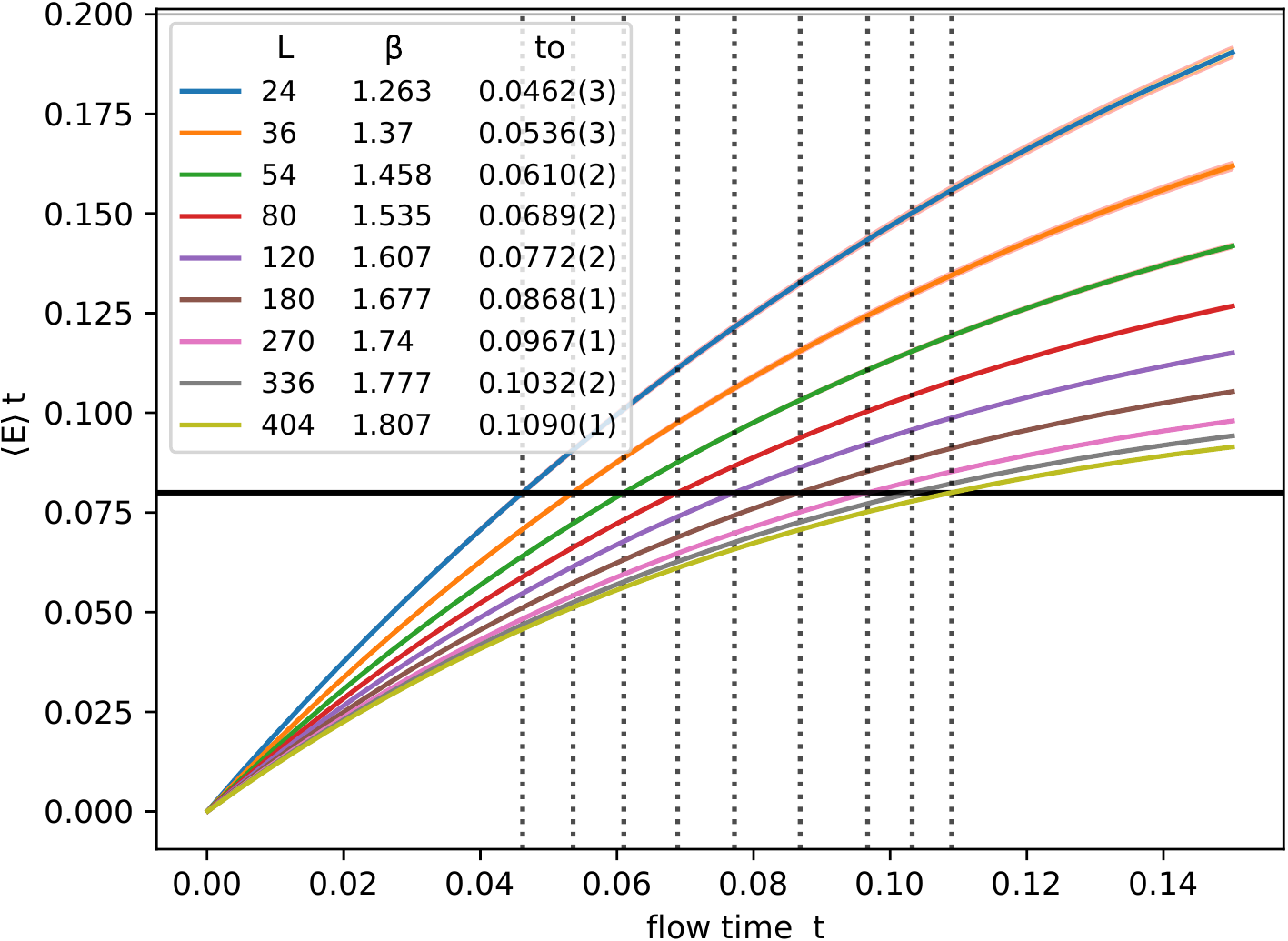}}\hfill
   \subfigure
 {\includegraphics[width=7cm,clip]{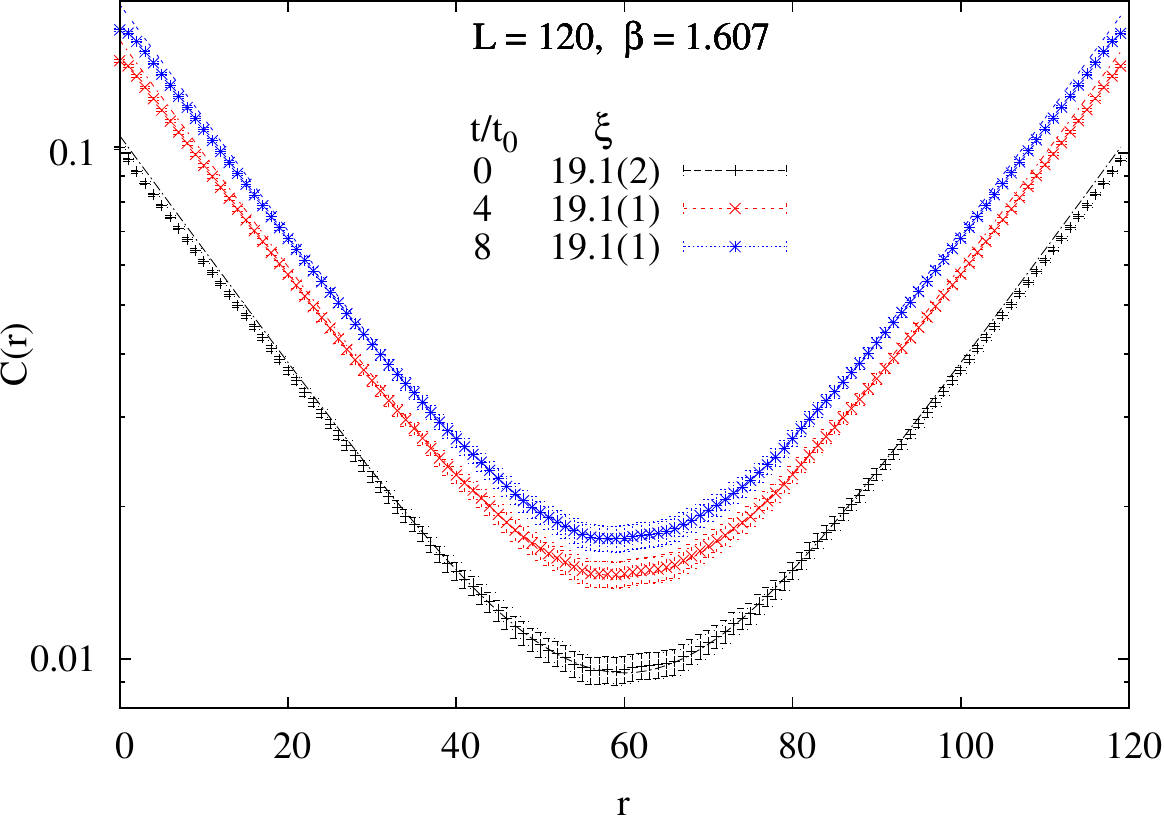}} 
 \vspace*{-2mm}
 \caption{The 2d O(3) model under GF: on the left, the term
   $\la E \ra t$, which fixes a GF time unit $t_{0}$ (in lattice units)
   by the condition
   $\la E \ra \, t_{0}= 0.08$. On the right: the correlation function
   $C(r)$ at $t= 0,\, 4t_{0}, \, 8t_{0}$. At fixed distance $r$ the
   GF moves it up, but the correlation length $\xi$ remains practically
   constant.}
\label{2dO3t0corr}
\vspace*{-5mm}
\end{figure}

Figure \ref{2dO3t0corr} (left) shows the early GF time evolution
of the product $\la E \ra t$. At longer flow times it increases
up to some maximum, before monotonically decreasing again.
For increasing volume, the value of the maximum of
$\la E \ra t$ decreases, hence the reference value has to be
sufficiently small, such that it is attained in all volumes
under consideration. We are in the process of extending this
study up to $L=494$ and $606$, where for instance $0.1$ is not
attained anymore. Hence we chose the reference value $0.08$, which
works up to $L=606$ \cite{prep}. So we refer to the definition
\be
\la E \ra \, t_{0} = 0.08 \ ,
\ee
as we anticipated in Figure \ref{2dO3slab}.

Figure \ref{2dO3t0corr} (right) shows an example for the
zero-momentum spin-spin correlation
function $C(r)$ at $t = 0,\, 4t_{0},\, 8t_{0}$. 
At a fixed distance $r$ between Euclidean time layers, $C(r)$
increases as the GF proceeds, but the correlation length $\xi$
remains unchanged within the errors. We measured $\xi$ by a fit to
a cosh-function in the range $r \in [L/3, 2L/3]$. We see that the
GF --- up to $t = 8 t_{0}$ --- hardly affects this long-range property.

\begin{figure}[tp]
   \centering
   \subfigure
 {\includegraphics[width=7cm,clip]{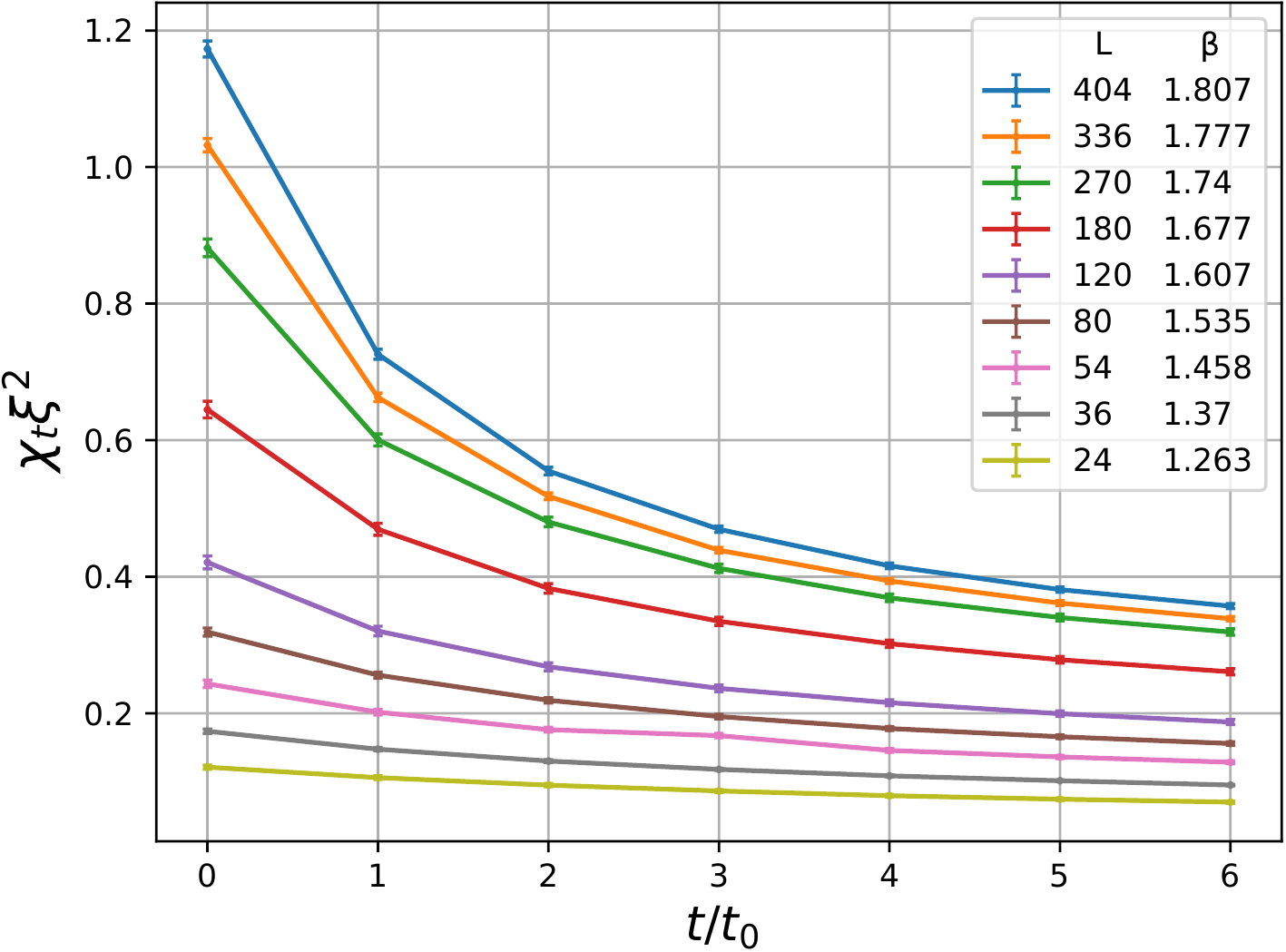}}\hfill
   \subfigure
 {\includegraphics[width=7cm,clip]{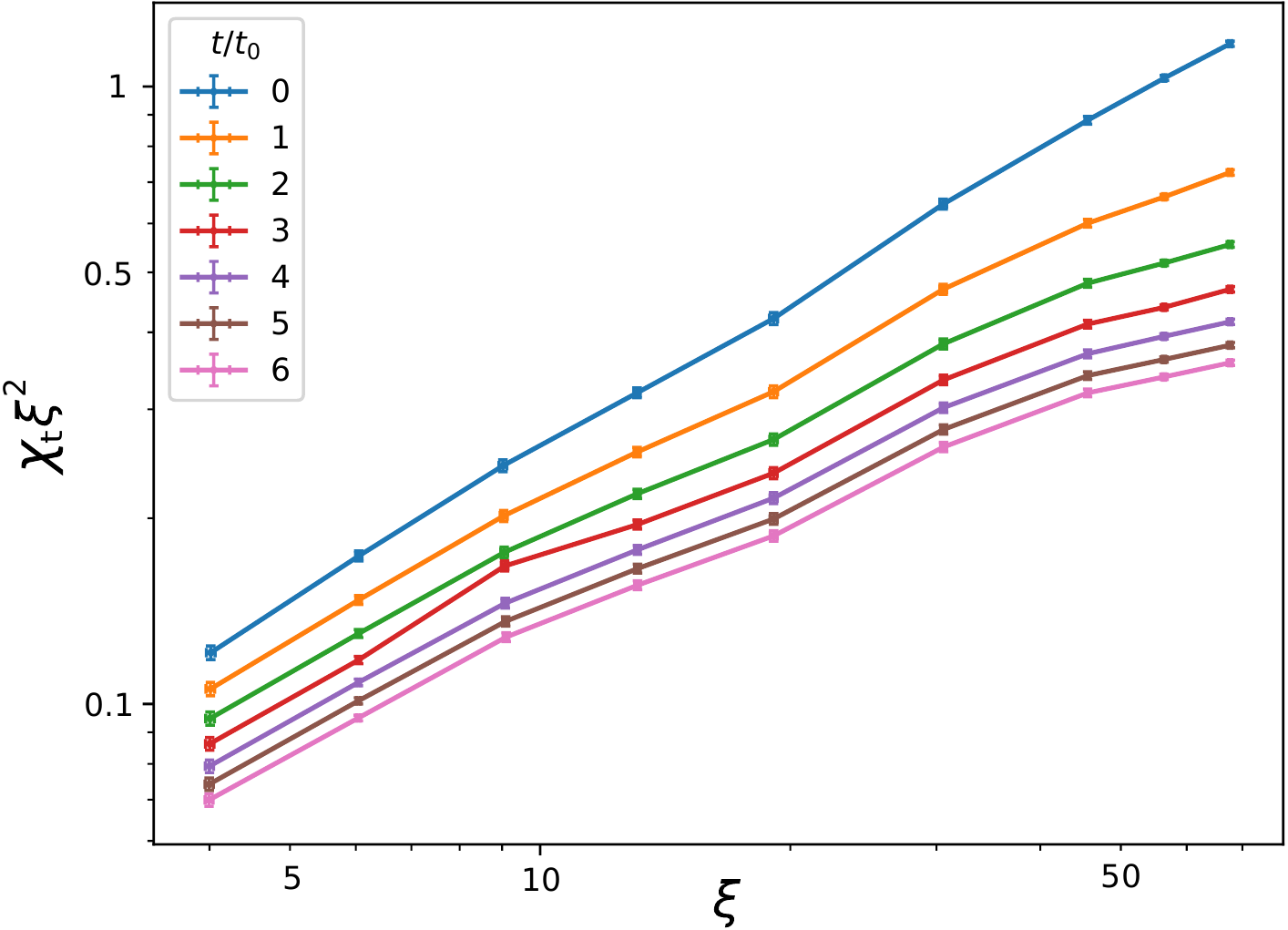}}
 \vspace*{-2mm}
 \caption{Left: the GF time evolution of the term $\chi_{\rm t} \xi^{2}$
   in nine $L \times L$ lattices, in each case at $L \simeq 6 \xi$. In
   large volumes we observe a rapid decrease at an early stage of the
   GF; this can be interpreted as the destruction of numerous small
   dislocations. Right: in the range $t=0 \dots 6 t_{0}$, the quantity
   $\chi_{\rm t} \xi^{2}$ does not seem to attain any finite continuum
   value as we approach the continuum (increasing $\xi$).}
\label{2dO3chit}
\vspace*{-5mm}
\end{figure}

Figure \ref{2dO3chit} (left) illustrates the GF time evolution
of the term $\chi_{\rm t} \xi^{2}$, which is supposed to be the
scaling quantity. We see a rapid decrease at an early stage of the
GF flow, in particular at large $L$ and $\xi$. This observation
is compatible with an increasing dominance of small dislocations
on fine lattices: these topological windings are destroyed even
by a short GF flow.

Finally, Figure \ref{2dO3chit} (right) shows $\chi_{\rm t} \xi^{2}$ as a
function of $\xi$, at various multiples of the flow time unit $t_{0}$
(the lines are drawn to guide the eye). At this stage, no trend towards
a stabilisation is visible. The observed behaviour at $t=0$ is well
compatible with a logarithmically divergent function of the
form $\chi_{\rm t} \xi^{2} = c_{1} \ln (c_{2} \xi + c_{3})$; at
$t = t_{0} \dots 6t_{0}$ the quality of this fit
is somewhat worse, but it still follows roughly this behaviour
(although a power-law $c_{1} \xi^{c_{2}} + c_{3}$
can be fitted with a similar quality) \cite{RA17}.
In any case, the present data do not provide a basis
for revising the standard lore of a topologically
ill-defined continuum limit in this model.

\section{Summary and outlook}\label{sumout}

In Section \ref{slab} we have discussed the slab method, which enables
a reliable measurement of the topological susceptibility within a
fixed topological sector, {\it i.e.}\ from a Markov chain at fixed
$Q$. This method does still work quite well when the GF is applied,
but in some cases --- in particular in QCD --- we observed the necessity
to subtract a constant from the expected fitting function.\\

In the 2d O(3) model, the data presented in Section \ref{2dO3}
do not suggest a continuum convergence of the term
$\chi_{\rm t} \xi^{2}$ after application of the GF.
Further details, including a table with numerical
results, are given in Ref.\ \cite{RA17}.

However, the impact range $\bar x (t)$ of the GF is short in these
examples. It can be estimated based on the heat kernel $K(t,x)$;
in $d$ dimensions we obtain
\be
K(t,x) = \frac{1}{(2\pi t)^{d/2}} \, \exp \Big( -x^{2}/(4t) \Big) \ ,
\quad \bar x (t) = \left( \int d^{d}x \, x^{2} K(t,x) \right)^{1/2}
  = \sqrt{2d \, t} \ .
\ee
In our 2d model it attains at most
$\bar x (6t_{0}) = \sqrt{24 t_{0}} \simeq 1.62$ at this stage
of our study; this refers to our largest volume, $L=404$,
with $\xi = 67.7(3) \gg \bar x (6t_{0})$.

In order to arrive at conclusive results, we are now going to fix
an extended GF time unit $T_{0}$ by a condition
$T_{0}/ \xi^{2} = {\rm constant}$, and investigate flow times up
to an impact range of $\bar x (t) \approx \xi/2$.
This study is in progress \cite{prep},
and it should finally reveal whether or not the GF leads
to a continuum scaling of the quantity $\chi_{\rm t} \xi^{2}$.\\

\noindent
{\bf Acknowledgements} \
We thank Martin L\"{u}scher for attracting our interest to the subject
of Section \ref{2dO3}, and for advice regarding the strategy towards
conclusive results. We further thank Marc Wagner for helpful discussions
about the slab method, which we discussed in Section \ref{slab}, and the
organisers of the 35th International Symposium on Lattice Field Theory.
This work was supported by DGAPA-UNAM, grant IN107915, by the
{\it Consejo Nacional de Ciencia y Tecnolog\'{\i}a} (CONACYT)
through project CB-2013/222812, and by the Helmholtz International
Center for FAIR within the framework of the LOEWE program launched
by the State of Hesse. K.C.\ was supported by the Deutsche
Forschungsgemeinschaft (DFG), project nr.\ CI 236/1-1, and A.D.\
by the Emmy Noether Programme of the DFG, grant WA 3000/1-1.

\bibliography{lattice2017}

\end{document}